\documentclass[runningheads]{svmult}

\usepackage{makeidx}   % allows index generation
\usepackage{graphicx}  % standard LaTeX graphics tool
\usepackage{subeqnar}  % subnumbers individual equations
\usepackage{multicol}  % used for the two-column index
\usepackage{physprbb}  % modified textarea for proceedings,
\makeindex             % used for the subject index
\begin{document}
\title*{Subaru Surveys for High-z Galaxies}
\toctitle{Subaru Surveys for High-z Galaxies}
% allows explicit linebreak for the table of content
%
%
\titlerunning{Subaru Surveys for High-z Galaxies}
% allows abbreviation of title, if the full title is too long
% to fit in the running head
%
\author{Yoshiaki Taniguchi}
\authorrunning{Yoshiaki Taniguchi}
% if there are more than two authors,
% please abbreviate author list for running head
%
%
\institute{Astronomical Institute, Graduate School of Science,
Tohoku University, Aramaki, Aoba, Sendai 980-8578, Japan}

\maketitle              % typesets the title of the contribution

\begin{abstract}
We present a summary of optical/NIR deep surveys for very high-$z$ galaxies
using the 8.2m Subaru Telescope operated by National Astronomical Observatory
of Japan. The prime focus mosaic CCD camera, Suprime-Cam, with a very wide
field of view, 34$^\prime \times 27^\prime$, allows us to carry out efficient 
optical deep surveys. In particular, the Subaru Deep Field project has
provided us a number of Lyman$\alpha$ emitters beyond $z=6$. 
We discuss the star formation history in the early universe
based on this project.
\end{abstract}

\section{Introduction}

Since the discovery of Ly$\alpha$ emission from a galaxy at $z=5.34$
\cite{dey98}, more than two dozen of Ly$\alpha$ emitters (LAEs)
have identified spectroscopically; see for reviews, \cite{tan03b};
\cite{spi03}. The most distant LAE known to date is SDF J132418.3+271455
at $z=6.578$ \cite{kod03}. Another very high-$z$ LAE is HCM-6A
at $z=6.56$ \cite{hu02}. These discoveries are actually thanks to
the great observational capability of 8-10m class optical telescopes.
Furthermore, the GOODS survey has provided a sample of very high-$z$
Lyman break galaxies (LBGs) at $z \sim 6$, thanks to the high-quality
imaging capability of the Advanced Camera for Surveys (ACS) on 
the Hubble Space Telescope (e.g., \cite{gia03}, \cite{dic03}, \cite{sta03}).
These exciting observations enable us to investigate the cosmic
star formation history and mass assembly history in the early universe.
In this review, we present a summary of recent deep surveys for
very high-$z$ (i.e., $z > 5$) galaxies based on the 8.2m Subaru Telescope.

\section{Subaru Surveys for High-$z$ Galaxies}
\subsection{The Subaru Deep Survey}

The 8.2m Subaru Telescope \cite{kai00} has seven instruments; 
see http://www.
subarutelescope.org/Observing/Instruments/index.html.
During the commissioning phase of three instruments (FOCAS, OHS/CISCO,
and Suprime-Cam), these instruments team members organized a systematic deep 
survey using these three instruments
to investigate high-$z$ galaxies; the Subaru Deep Survey (SDS).
All the observations were done during a period between 1999 and 2001.
Their target fields are (1) the Subaru Deep Field (SDF) centered  at
RA(J2000) = 13$^{\rm h}$ 24$^{\rm m}$ 21.$^{\rm s}$38 and 
DEC(J2000) = $+ 27^\circ 29^\prime 23^{\prime\prime}$, and (2)
the Subaru XMM-Newton Deep Field (SXDF) centered  at
RA(J2000) = 2$^{\rm h}$ 18$^{\rm m}$ 00.$^{\rm s}$00 and 
DEC(J2000) = $- 5^\circ 12^\prime 00^{\prime\prime}$.
The SDF is used to make a very deep imaging survey while the SXDF
is used to make a wide-field, medium deep one; see for the SXDF
project, \cite{sek03}, \& \cite{ouc03b}.

(1) NIR Deep Imaging Survey: Very deep $J$ and $K^\prime$ images 
of the central 2$^\prime \times 2^\prime$ field of the SDF were 
obtained with use of CISCO \cite{mai01}.
The integration times
of the  $J$ and $K^\prime$ bands were 12.1 hr and 9.7 hr, resulting
in 5$\sigma$ limiting magnitudes of 25.1 and 23.5 mag (the Vega system),
respectively. These data are used to investigate the NIR galaxy number
count, colors, and size distribution; see also \cite{tot01c}.
They also found a population of hyper extremely red objects (HEROs) with
$J - K^\prime >$ 3 -- 4 \cite{tot01b}. These deep NIR data 
were also utilized to investigate the diffuse extragalactic background 
light (EBL) \cite{tot01a}. They found that $\sim$ 90 \% of the EBL 
from galaxies were resolved in their deep NIR images.

This NIR data set was also used to construct a $K^\prime$-selected 
galaxy sample, consisting of 439 galaxies for which both
optical ($B$, $V$, $R$, $I$, and $z^\prime$)  and NIR photometric data 
are available \cite{kas03}. Comparing the star
formation rate density (SFRD) at $z \sim 3$ for their $K^\prime$-selected 
sample with those based on previous LBG surveys, they found that 
a large fraction of SFRD at $z>1.5$ may come from a faint blue galaxy
population.

(2) Optical Narrowband Deep Survey: One of narrowband filters, NB711 
centered at $\lambda_{\rm C} = 7126$ \AA ~ with $\Delta \lambda$ = 73 \AA ~
was used to search for LAEs at $z \sim 4.9$ \cite{ouc03a}, \cite{shi03}.
They found 87 reliable LAE candidates
at $z \sim 4.9$, and then analyzed their luminosity function
and clustering properties \cite{ouc03a}. They also found a
large-scale clustering of LAEs with a scale of $\sim$ 20 Mpc $\times$
50 Mpc \cite{shi03}.

(3) Optical Broad Band Deep Survey: In order to investigate  
photometric and clustering properties of LBGs at $z \sim$ 4 -- 5,
optical broad band data of both the SDF and the SXDF,
covering 1200 sq. arcmin in total were carefully analyzed by
\cite{ouc03b}, \cite{ouc03c}. They obtained a large
sample of LBGs (2600 objects) at $z \simeq$ 3.5 -- 5.2.

Their analysis shows that the correlation lengths
are $\simeq $ 4.1 $h_{100}^{-1}$ Mpc and 5.9  $h_{100}^{-1}$ Mpc
in co-moving units for all the detected LBGs at $z \simeq 4$
and $z \simeq 5$, respectively.They also found that 
a typical mass of dark matter halos hosting LBGs with $L > L^*$
amounts to $\sim 1 \times 10^{12} M_\odot$, being comparable
to those of typical massive disk galaxies like our Milky Way.

Based on the CDM model, they also estimated the mass of dark matter 
halos which could form from such high-$z$ objects. Since they
obtained a mass range between $\sim 10^{13}$ -- $10^{15} M_\odot$,
they suggested that dark matter halos hosting
high-$z$ LBGs could evolve to groups and clusters in the local universe.
On one hand, faint LBGs, LAEs, and $K^\prime$-selected galaxies
could evolve to present-day galaxies after experiencing a few
merger events.

\subsection{The Subaru Deep Field (SDF) Project}

As outlined in the previous subsection, the SDS gave a number of 
important findings in the research field of galaxy evolution.
This success seems to be attributed to the very wide-field of view
of Suprime-Cam and excellent seeing conditions at Mauna Kea.
In order to make the SDS much more fruitful, the Subaru Telescope Office 
decided to promote big surveys using guaranteed observing time that each 
Subaru builder member has. Then they proposed two big surveys for the 
extragalactic research; (i) the Subaru Deep Field Project led by Nobunari 
Kashikawa, and (ii) the Subaru XMM-Newton Deep Survey Project led by 
Kaz Sekiguchi. As mentioned before, the latter project is aimed to carry 
out a wide-field (1 sq. degree), medium deep survey in collaboration
with the XMM-Newton Observatory. Since the SDF is dedicated to
 a very deep search for high-$z$ galaxies, we present a brief summary of the 
current status of the SDF.
It is noted that a common-use, intensive program on ^^ ^^ A Search
 for Ly$\alpha$ Emitters at $z=5.7$ and $z=6.6$" (Proposal ID = S02A-IP2; 
PI = Y. Taniguchi) joined to the SDF project.

(1) SDF2002: Thirteen nights were allocated to the SDF project 
in the semester S02A. In this semester, we performed a deep optical 
imaging survey using a narrowband filter ($NB921$) centered at
 $\lambda =$ 9196 \AA ~ together with $i^\prime$
and $z^\prime$  broadband filters covering an 814 arcmin$^2$ area of the SDF.
We obtained a sample of 73 strong  $NB921$-excess objects
based on the following two color criteria;
$z^\prime - NB921 > 1$ and $i^\prime - z^\prime > 1.3$.
We then obtained optical spectroscopy of nine objects
in our $NB921$-excess sample, and identified at least two
Ly$\alpha$ emitters at$z=6.541 \pm 0.002$ and $z=6.578 \pm 0.002$, 
each of which shows the characteristic sharp cutoff
together with the continuum depression at wavelengths
shortward of the line peak.

These new data allow us to estimate the first meaningful lower limit of
the star formation rate density beyond redshift 6 \cite{kod03}.
First, we estimate the total star formation rate of 73 LAEs in our 
photometric sample using the equivalent width of NB921 flux.
Our follow-up optical spectroscopy found that two among the nine 
LAE candidates are real LAEs, it seems reasonable to assume that
approximately 22\% (=2/9) of
73 LAE candidates are real LAEs at $z \approx$ 6.5 - 6.6;
$f({\rm LAE}) \simeq 22$\%.
If we assume that all the 73 LAE candidates are true LAEs 
at $z \approx 6.5$ -- 6.6,
we obtain nominally a total star formation rate of 
$SFR_{\rm total}^{\rm nominal} = 475 h_{0.7}^{-2} ~ M_\odot$ yr$^{-1}$.
Adopting $f({\rm LAE}) \simeq 22$\%,
we can estimate the total star formation rate,
$SFR_{\rm total} \simeq 0.22 \times SFR_{\rm total}^{\rm nominal} 
\simeq 105  h_{0.7}^{-2} ~ M_\odot$ yr$^{-1}$.
Given the survey volume, 202,000 $h_{0.7}^{-3}$ Mpc$^{3}$, we thus
obtain a star formation rate density of $\rho_{\rm SFR} \simeq 5.2 \times 10^{-4}
h_{0.7} ~ M_\odot$ yr$^{-1}$  Mpc$^{-3}$.
This observation reveals
that a moderately high level of star formation activity
already occurred at $z \sim$ 6.6 (see also \cite{hu02}).

(2) SDF2003: Fifteen nights were allocated for the SDF project 
in the semester S03A. We made optical deep imaging and spectroscopy again,
and finished our optical imaging survey. 
We obtained optical spectra of additional 18 LAE candidates using FOCAS,
and thus we obtained a spectroscopic sample of 27 LAE candidates including 
our spectroscopy made in 2002. From our spectroscopy, we identify
nine LAEs at $z=6.50$ -- 6.60. The remaining 18 objects are;
nine single-line emitters, one [O {\sc ii}] emitter at $z=1.46$,
two [O {\sc iii}] emitters at $z=0.84$ and $z=0.85$, and six
unclassified objects. The single-line emitters are either [O {\sc ii}] emitters
at $z \sim1.46$ or LAEs at $z \sim 6.6$. Much higher-resolution
spectroscopy will be necessary to identify them unambiguously.

Since our new spectroscopy leads to a new value of $f({\rm LAE}) = 9/27 \simeq 
33$\%, we obtain a star formation rate density of $\rho_{\rm SFR} \simeq 
7.8 \times 10^{-4} h_{0.7} ~ M_\odot$ yr$^{-1}$  Mpc$^{-3}$.
It should be reminded that
we apply neither any reddening correction nor integration by assuming
a certain luminosity function for LAEs. Therefore, this value
should be regarded as a lower limit.

We also made follow-up spectroscopy of a small sample of
very red objects in $i^\prime - z^\prime$ color, and then
identified a new bright LAE at $z = 6.33$. 
This suggests that a number of LAEs may be found in such 
very red objects. 
Finally, we remind you that the data reduction of SDF data taken in 2003
is still underway.  

\begin{figure}[b]
\begin{center}
\includegraphics[width=.5\textwidth]{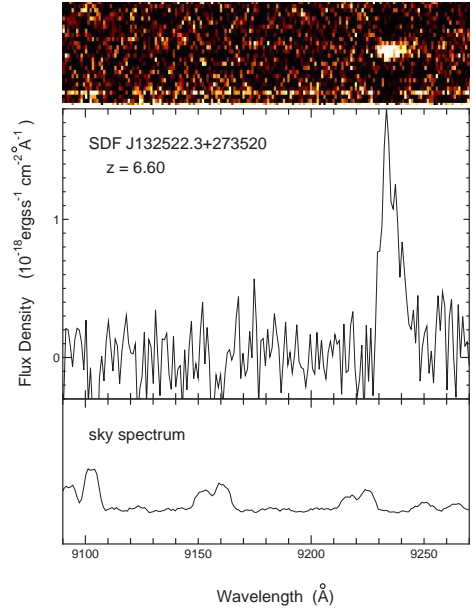}
\end{center}
\caption[]{The most distant LAE found in the SDF project;
SDF J132522.3+273529 at $z = 6.60$.}
\label{eps1}
\end{figure}

\subsection{Deep Surveys based on Common-Use Observations}

(1) Lyman Break Galaxies at $z \sim 5$: \cite{iwa03}
(Proposal ID = S00-017; PI = K. Ohta) made
deep optical imaging of 618 arcmin$^2$ including the Hubble Deep
Field-North to search for LBGs at $z \sim 5$. They found $\sim$
100 LBG candidates at $23.0 \leq I_{\rm C} \leq 24.5$ and $\sim$
300 LBG candidates at $23.0 \leq I_{\rm C} \leq 25.5$. These data
were used to estimate the rest-frame UV luminosity function 
at $4.4 \leq z \leq 5.3$. They found that the UV luminosity
density at this redshift range is lower by a factor of two than
that at $z \sim 3$.

(2) Lyman$\alpha$ Emitters at $z > 5$: \cite{aji03}
(Proposal ID =  S01B-051; PI = Y. Taniguchi) made
a survey for Ly$\alpha$ emitters at $z\approx 5.7$ based on optical 
narrow-band ($\lambda_{\rm c} = 8150$ \AA ~ and
$\Delta\lambda = 120$ \AA), and broad-band ($B$, $R_{\rm C}$, $I_{\rm C}$, 
and $z^\prime$) observations of the field surrounding the high
redshift quasar, SDSSp J104433.04$-$012522.2 at $z=5.74$.
This survey covers a sky area of $\approx 720$ arcmin$^2$ and a co-moving
volume of $\simeq 2 \times 10^5 ~ h_{0.7}^{-3}$ Mpc$^3$. They found 20 LAE
candidates at $z \approx$ 5.7 with $\Delta z \approx 0.1$.
This survey leads to a new estimate of the star formation rate density at 
$z \approx 5.7$, $\sim 1.2 \times 10^{-3} h_{0.7} M_\odot$ yr$^{-1}$  Mpc$^{-3}$.
It is also noted that this NB816 survey was used to investigate 
field H$\alpha$ emitters at $z \approx 0.24$ \cite{fuj03b}.

Among their 20 LAE candidates, two objects were confirmed star-forming galaxies at
$z=5.655$ and $z=5.687$ from their follow-up optical spectroscopy made with
FOCAS on Subaru and/or ESI on Keck II. 
LAE J1044$-$0130 is identified as a probable superwind galaxy at  
$z=5.687 \pm 0.002$ \cite{aji03}.
Its emission line profile is strongly truncated at
wavelengths blueward shortward the line peak while shows red-wing emission.
The observed broad line width, FWHM (full width at half maximum)
$\simeq$ 340 km s$^{-1}$ as well as the red wing emission suggest
that this object is experiencing the superwind activity.
The emission-line morphology appears to show a triangle shape.
This may be also interpreted in terms of the superwind activity.

LAE J1044$-$0123 is identified as a star forming galaxy at
$z=5.655\pm0.002$ with a star formation rate of $\sim 13 ~ h_{0.7}^{-2} ~ M_\odot$
yr$^{-1}$ \cite{tan03a}.
Remarkably, the velocity dispersion of Ly$\alpha$-emitting
gas is only 22 km s$^{-1}$.  Since a blue half of the Ly$\alpha$ emission
could be absorbed by neutral hydrogen gas, perhaps in the system,
a modest estimate of the velocity dispersion may be $\sim$ 44 km s$^{-1}$.
Together with a linear size of 7.7 $h_{0.7}^{-1}$ kpc,
we estimate a lower limit of
the dynamical mass of this object to be $\sim 2 \times
10^9 M_\odot$.
Therefore, LAE J1044$-$0123 seems to be a star-forming
dwarf galaxy (i.e., a subgalactic object or a building block).

\cite{aji04} also made a unique deep survey for LAEs at 
$z \sim 5.8$ using an intermediate-band filter
centered at $\lambda_{\rm c} \approx $ 8270 \AA ~ 
with $\Delta\lambda_{\rm FWHM} \approx $ 340 \AA ~
(i.e., the spectroscopic resolution is $R \approx 23$)
during the same observing run as that of \cite{aji03};
see for details of this intermediate-band filter system \cite{tan01},
In this survey,
they found four Ly$\alpha$-emitter candidates from the intermediate-band image
($z \approx$ 5.8 with $\Delta z \approx 0.3$);
see also \cite{fuj03a} for a similar survey for LAEs at $z \simeq 3.7$
using another intermediate-band filter IA 574.

In the above LAE survey, they observed a sky are surrounding
the high redshift quasar, SDSSp J104433.04$-$012522.2 at $z=5.74$.
They found a foreground lensing galaxy with
$m_B$(AB) $\approx 25$, located at 1.9 arcsec southwest of the quasar
\cite{shi02}.
Its broad band color properties from $B$ to $z^\prime$ suggest that
the galaxy is located at a redshift of $z \sim 1.5$ - 2.5.
Since the counter
image of the quasar cannot be seen in our deep optical images, the
magnification factor seems not so high. Our modest estimate is that
this quasar is gravitationally magnified by a factor of 2
;see also \cite{yam03}.

\section{Concluding Remarks}

The Hubble Space Telescope and 8-10m class optical telescopes 
have been contributing to the progress in deep searches for high-$z$ galaxies.
Although the Subaru Telescope came later to this research field,
as we see above, it is also powerful to search for high-$z$ galaxies
as well as the other 8-10m class telescopes.

Up to date, approximately several tens of LAEs beyond $z=5$ have 
already been identified spectroscopically. However, we still need 
any systematic deep surveys for such LAEs to understand the whole
history of cosmic star formation in the early universe. 
In particular, one of important things related to LAEs is 
to construct reliable Ly$\alpha$ luminosity functions of LAEs
as a function redshift and the compare them UV luminosity functions;
see for recent progress, \cite{aji03}, \cite{hu03}, \& \cite{san03}.

%\vskip{0.5cm}

We would like to thank all the SOC and LOC members, in particular,
Alvio Renzini and Ralf Bender.
We would like to thank Keiichi Kodaira, Norio Kaifu, Hiroyasu Ando, Hiroshi 
Karoji, Msanori Iye, Hy Spinrad,
Nobunari Kashikawa, Yutaka Komiyama, Sadanori Okamura, Kazuhiro Shimasaku, 
Masami Ouchi, Yasuhiro Shioya, Takashi Murayama, and Tohru Nagao for
useful discussion and encouragement.
We also thank all members of the SDF project.

\end{document}